# A Computational Model of Systems Memory Consolidation and Reconsolidation

Peter Helfer[1] and Thomas R. Shultz[1]

[1]McGill University, 2001 McGill College, Montreal, QC, Canada H3A 1G1

Acknowledgements

This work was supported by a Discovery Grant to TRS from the Natural Sciences and Engineering Research Council of Canada (7927-2012-RGPIN). The funding source had no other role than financial support.

Abstract

In the mammalian brain, newly acquired memories depend on the hippocampus for maintenance and recall, but over time the neocortex takes over these functions, rendering memories hippocampus-independent. The process responsible for this transformation is called *systems memory consolidation*. However, reactivation of a well-consolidated memory can trigger a temporary return to a hippocampus-dependent state, a phenomenon known as *systems memory reconsolidation*. The neural mechanisms underlying systems memory consolidation and reconsolidation are not well understood. Here, we propose a neural model based on well-documented mechanisms of synaptic plasticity and stability and describe a computational implementation that demonstrates the model's ability to account for a range of findings from the systems consolidation and reconsolidation literature. We derive several predictions from the computational model, and suggest experiments that may test its validity.

*Keywords:* memory reconsolidation; artificial neural network; AMPA receptor exchange; neural plasticity; computational model

# A Computational Model of Systems Memory Consolidation and Reconsolidation

The neural processes that transform memories from short-term to long-term storage are collectively known as *memory consolidation*. They include *synaptic consolidation*, relatively rapid intra-cellular changes that stabilize synaptic potentiation, and *systems consolidation*, slower and larger-scale processes that reorganize and restructure memory traces across brain systems. Specifically, systems consolidation refers to mechanisms that gradually make memories independent of the hippocampus, a structure in the medial temporal lobe of the mammalian brain.

Whereas new memories are susceptible to disruption by a number of different types of interventions (e.g. electroconvulsive shock, certain pharmaceuticals, surgical procedures, and interference from new learning), consolidated memories are resistant to these treatments. However, retrieval of a consolidated memory can trigger a process in which it transiently becomes vulnerable to such interventions again, but subsequently restabilizes into a consolidated state. This is known as *reconsolidation*, and like consolidation it can be observed both at the synaptic and systems level. While much has been learned about the molecular underpinnings of synaptic consolidation and reconsolidation, the mechanisms responsible for the systems-level phenomena remain largely unknown (Asok, Leroy, Rayman, & Kandel, 2019; Hardt & Nadel, 2018).

Here, we present an artificial neural network model that includes connection dynamics based on mechanisms of synaptic plasticity and demonstrate how these low-level processes can account for systems consolidation and reconsolidation.

We begin with overviews of synaptic and systems memory consolidation and reconsolidation, and of previously published computer simulations. Next, we describe our model, report on simulation results and discuss their implications.

## Synaptic Consolidation and Reconsolidation

**Synaptic transmission.** Neurons generate electrical signals called action potentials (APs) that travel along nerve fibers (axons) toward synapses where connections are made with other neurons. When an action potential reaches a synapse, neurotransmitter is released into the *synaptic cleft*, a narrow gap between the presynaptic *active zone* and the *post-synaptic density* (*PSD)*, specialized areas of neuronal cell membrane that together make up the synapse. The neurotransmitter molecules bind to receptor proteins in the PSD, thereby triggering a response in the postsynaptic neuron. Depending on the type of neurotransmitter and the type of receptor, the response may be excitatory or inhibitory (making the postsynaptic neuron more or less likely to generate an AP), or have some other, e.g. regulatory, function. The average size of the excitatory or inhibitory response that is generated by the arrival of an action potential at a particular synapse is a measure of synaptic strength, and it depends both on the amount of transmitter released and on the numbers and types of receptors in the PSD (Citri & Malenka, 2008; Kandel, Dudai, & Mayford, 2014). Most neuroscientists believe that memories are stored in the strengths of synapses (Kandel et al., 2014; Sossin, 2008), an idea first articulated by Santiago Ramón y Cajal in the late 19$^{th}$ century (1894) and known as *the synaptic theory of memory* (Langille & Brown, 2018).

**Glutamate receptors.** The amino acid glutamate is the most abundant neurotransmitter in the vertebrate nervous system (Platt, 2007). There are several types of glutamate receptors, among which the α-amino-3-hydroxy-5-methyl-4-isoxazolepropionic acid receptor (AMPA receptor or AMPAR) is chiefly responsible for mediating excitatory synaptic transmission (Citri & Malenka, 2008; Maren, Tocco, Standley, Baudry, & Thompson, 1993), and the N-methyl-D-aspartate receptor (NMDA receptor or NMDAR) is involved with regulatory functions including the regulation of synaptic strength (Ben Mamou, Gamache, & Nader, 2006; Citri & Malenka, 2008; D. L. Walker & Davis, 2002).

**Long-term potentiation**. When a neuron is stimulated strongly enough to make it fire (generate an AP), participating glutamatergic synapses are strengthened by a process called long-term potentiation (LTP) (Bliss & Lømo, 1973), which is associated with an increase in the number of AMPARs inserted in the PSD (Malenka & Bear, 2004). There are different stages of LTP. Moderately strong stimulation gives rise to early-phase LTP (E-LTP), which lasts for at most a few hours. More intense stimulation can trigger the induction of late-phase LTP (L-LTP), which can persist for months or longer (Abraham, 2003).

Many researchers believe that LTP as studied in the laboratory is a cellular model for learning and memory, i.e. that it replicates the synaptic changes that occur during memory formation (Abraham, 2003; Bliss & Collingridge, 1993; Kandel, 2000; Lisman, 2017; Pastalkova et al., 2006), although conclusive proof for this has proved difficult to obtain (Lynch, 2004; Stevens, 1998).

**Consolidation**. Induction of L-LTP – but not E-LTP – is believed to require RNA translation (synthesis of new proteins), based on experiments with protein synthesis inhibiting drugs (PSIs) such as anisomycin or cycloheximide. Infusion of such drugs before or immediately after stimulation can prevent establishment of L-LTP (Fonseca, Nägerl, & Bonhoeffer, 2006; Frey, Huang, & Kandel, 1993; Frey, Krug, Reymann, & Matthies, 1988; Huang, Li, & Kandel, 1994). However, once L-LTP has been established, a process that takes on the order of one hour, it is no longer vulnerable to PSI infusion (Bourtchouladze et al., 1998; Fonseca et al., 2006; Kandel, 2000).

At the behavioral level, PSI injection within the first hour after training, but not later, has been shown to cause memory impairments (Davis & Squire, 1984; Judge & Quartermain, 1982; R. Kim, Moki, & Kida, 2011; Lattal & Abel, 2004), suggesting that the formation of long-term memory similarly requires protein synthesis (but see Canal and Gold (Canal, Chang, & Gold, 2007), who argue that, at least in the case of anisomycin, the memory impairments may be caused by other effects of the drug, not by protein synthesis inhibition).

**Reconsolidation**. Several studies from the 1940s and 1950s demonstrated that electroconvulsive shock (ECS) could interfere with the establishment of long-term memory in rodents (Duncan, 1949; Gerard, 1955; Thompson & Dean, 1955) and in humans (Cronholm & Lagergren, 1959; Kehlet & Lunn, 1951), but only when applied within an hour or two after acquisition. However, in 1968, Misanin et al. reported that ECS could impair 24-hour-old, i.e. consolidated, memories of fear conditioning in rats - but only if the convulsive treatment was immediately preceded by memory “reactivation”, i.e. retrieval cued by presentation of the conditioned stimulus (Misanin, Miller, & Lewis, 1968). Post-reactivation susceptibility to ECS was also demonstrated by Schneider and Sherman (1968) and Lewis, Mahan and Bregman (1972). Judge and Quartermain (1982) reported that injection of the protein synthesis inhibitor anisomycin, which was known to produce memory deficits when administered to mice immediately after training, could also impair older memories if given 30 minutes or less after reactivation. Przybyslawski and Sara (1997) showed that the NMDA receptor antagonist MK-801 could induce memory deficits in rats that had been trained on a maze-running task, if injected up to 90 minutes after a maze run, but not after 120 minutes. The authors proposed that reactivation returns a well-established memory to a labile state from which it normally restabilizes spontaneously, and that this restabilization requires some or all of the same NMDA receptor-dependent events that are needed for consolidation of new memories. They therefore referred to the process as *memory reconsolidation* (Przybyslawski & Sara, 1997), a term first introduced by Spear (1973). Nader et al. (2000) demonstrated that anisomycin infusion into the amygdala of rats could disrupt an established fear-conditioning memory if performed immediately after reactivation, but not six hours later. Taken together, these studies support the notion that reactivation can render a memory trace that has undergone synaptic consolidation labile, and that an NMDA-dependent process, likely involving protein synthesis, is required to subsequently restabilize it. The phenomenon, known as *synaptic memory reconsolidation,* has attracted much interest in the wake of the Przybyslawski and Sara (1997) and Nader et al. (2000) papers, and a large reconsolidation literature now exists (for reviews, see Baldi & Bucherelli, 2015; Besnard, Caboche, & Laroche, 2012; Lee, Nader, & Schiller, 2017; Nader & Einarsson, 2010).

A reconsolidation-like phenomenon has also been observed at the synaptic level: protein synthesis inhibition normally does not disrupt established (i.e. several hours old) L-LTP, but if administered together with low-frequency electric stimulation, it can cause depotentiation (Fonseca et al., 2006; Okubo-Suzuki et al., 2016).

**The function of reconsolidation.** It is generally believed that the function of post-retrieval plasticity is to permit memory modification or updating when new information is encountered (Lee, 2009; Lee et al., 2017; Schiller & Phelps, 2011). Several studies with human subjects have shown that new training material is more likely to interfere with an established memory if presented after reactivation of the original memory (Forcato et al., 2007; Hupbach,

Gomez, Hardt, & Nadel, 2007; M. P. Walker, Brakefield, Hobson, & Stickgold, 2003), although some authors have questioned whether such results really are evidence of a reconsolidation process (Ecker & Lewandowsky, 2012; Sederberg, Gershman, Polyn, & Norman, 2011). Post-reactivation memory updating has also been demonstrated in rodents (Jones, Pest, Vargas, Glisky, & Fellous, 2015; Monfils, Cowansage, Klann, & LeDoux, 2009).

**The molecular underpinnings of LTP.** The biochemical basis of LTP is not completely understood, but intense research efforts during the last several decades have begun to throw light on some of the underlying molecular mechanisms (Baltaci, Mogulkoc, & Baltaci, 2019; Kandel et al., 2014). One significant discovery is that different types of AMPA receptors are of importance for the induction and maintenance of the early and late phases of LTP (Hong et al., 2013; McCormack, Stornetta, & Zhu, 2006; Rao-Ruiz et al., 2011). An AMPA receptor is made up of four subunits, each of which can be of several different kinds. Depending on its subunit composition, an AMPA receptor may or may not permit calcium ions to pass through the cell membrane, and is accordingly designated as calcium-permeable (CP) or calcium-impermeable (CI) (Henley & Wilkinson, 2013). E-LTP induction is characterized by a rapid increase in the number of CP-AMPARs, while the establishment of L-LTP requires insertion of CI-AMPARs (Clem & Huganir, 2010; Hong et al., 2013; Kessels & Malinow, 2009; Plant et al., 2006). This discovery is significant because the increased CP-AMPAR count is relatively short-lived, whereas an elevated level of CI-AMPARs can be sustained for a long time (Clem & Huganir, 2010; Hong et al., 2013; Plant et al., 2006). The ability of CI-AMPARs to remain at the PSD for a long time has been hypothesized to be due to molecular mechanisms that traffic CI-AMPARs to the potentiated synapse and protect them against removal (Clopath, Ziegler, Vasilaki, Buesing, & Gerstner, 2008; Helfer & Shultz, 2018; Jalil, Sacktor, & Shouval, 2015; Sacktor, 2011;

Smolen, Baxter, & Byrne, 2012; Zhang, Smolen, Baxter, & Byrne, 2010). Interestingly, memory retrieval has been shown to trigger a transient reversal to a state with high CP-AMPAR and low CI-AMPAR counts (Hong et al., 2013; Rao-Ruiz et al., 2011), providing a potential explanation for post-retrieval synaptic instability. As discussed above, infusion of a protein synthesis inhibitor can interfere both with the induction of L-LTP and with restabilization of L-LTP after reactivation. A proposed explanation for this is that proteins required to transport CI-AMPARs into the synapse, and to maintain them there, need to be synthesized in order for either of these processes to occur (Sacktor, 2011).

Additional evidence supporting the notion of LTP as a memory model includes other pharmacological interventions that affect both LTP and memory. For example, inhibition of NMDA receptors blocks both LTP and fear memory acquisition (Fanselow & Kim, 1994; Miserendino, Sananes, Melia, & Davis, 1990), inhibition of the protein kinase PKMζ with ZIP (zeta-inhibitory peptide) disrupts both L-LTP (Ling et al., 2002; Serrano, Yao, & Sacktor, 2005) and long-term memory (LTM) (Pastalkova et al., 2006; Shema, Hazvi, Sacktor, & Dudai, 2009; Shema, Sacktor, & Dudai, 2007), and blocking endocytosis (removal from the cell membrane) of CI-AMPARs rescues both L-LTP and LTM from these effects of ZIP (Migues et al., 2010, 2016). Also, optogenetic stimulation protocols that induce long-term depression (LTD) and LTP have been shown to inactivate and reactivate, respectively, a conditioned fear response (Nabavi et al., 2014).

To summarize, LTP is considered a cellular model of memory, because it is compatible with the synaptic theory of memory, and although a causal relationship has proved difficult to establish, many results support the notion that the synaptic changes that characterize LTP also play an important role in the cellular mechanisms of memory.

## Systems Consolidation and Reconsolidation

**Consolidation.** Scoville and Milner (1957) famously documented that bilateral hippocampal lesions in human patients resulted in profound memory loss for past events (retrograde amnesia), as well as a nearly complete inability to form new long-term memories (dense anterograde amnesia). The retrograde amnesia appeared to be graded: it was most severe for the period shortly before surgery, while older memories were relatively spared. These findings have been confirmed with other human patients (Dede & Smith, 2016; Manns, Hopkins, & Squire, 2003; Penfield & Milner, 1958), and reproduced in animal studies with primates (Squire & Zola-Morgan, 1991; Zola-Morgan & Squire, 1990) and rodents (Anagnostaras, Maren, & Fanselow, 1999; Winocur, 1990) (although a temporal gradient is not always observed (Lah & Miller, 2008; Sutherland, Sparks, & Lehmann, 2010)). The observation that recent memories are more vulnerable to hippocampal damage than older ones has given rise to the notion that the hippocampus plays a crucial role in the establishment, maintenance and recall of new memories, but over time memories become hippocampus-independent (Frankland & Bontempi, 2005; McClelland, McNaughton, & O'Reilly, 1995; Milner, 1989; Nadel & Hardt, 2010; Squire & Zola-Morgan, 1991). The process responsible for this putative reorganization is called *systems memory consolidation* (Dudai, 2004). According to what is called *the standard theory of systems consolidation* (Squire & Alvarez, 1995), the hippocampus quickly records information about events in real time, then repeatedly replays them, perhaps primarily during sleep (Genzel et al., 2017), thereby driving a more time-consuming process of memory trace creation in the neocortex (Frankland & Bontempi, 2005; McClelland et al., 1995; Teyler & Rudy, 2007). A possible explanation for why it takes longer to lay down a memory trace in neocortex is that it is more sparsely inter-connected than the hippocampus, and that creation of a memory trace therefore requires axonal growth and synaptogenesis (Alvarez & Squire, 1994; Frankland & Bontempi, 2005). Another possibility, suggested by Teyler and Rudy (2007), is that strong inhibitory control mechanisms in the neocortex slow down the induction of LTP.

The idea that hippocampal "replay" can trigger reinstatement of activation patterns in neocortex raises the question of what kind of information is stored in the hippocampus. McClelland et al., in their "two memory systems" proposal (1995), suggested that the hippocampus stores "compressed representations" of cortical patterns, that can be decoded to reconstruct activation patterns in neocortex, but most authors favor a view where the hippocampus does not store memory content, but rather "indices" (Teyler & Discenna, 1986; Teyler & Rudy, 2007) or "links" to loci in the neocortex where components of memory traces reside (Alvarez & Squire, 1994; Frankland & Bontempi, 2005). In these accounts, the hippocampus plays a dual role: it supports, through replay, the establishment and/or strengthening of intra-neocortical connections, and it also assist in memory retrieval until those connections become strong enough to function independently.

Not all researchers agree with the standard theory of systems consolidation. In particular, proponents of the Multiple-Trace Theory (Nadel & Moscovitch, 1997) and its descendant the Trace Transformation Hypothesis (Moscovitch, Cabeza, Winocur, & Nadel, 2016; Nadel, Winocur, Ryan, & Moscovitch, 2007; Winocur, Sekeres, Binns, & Moscovitch, 2013) hold that whereas semantic memories (knowledge about the world) are consolidated in the manner described by the standard theory, autobiographical memories (spatial and temporal information) remain hippocampus-dependent for as long as they persist. According to this view, the loss of hippocampal support (whether due to lesion/damage or to gradual weakening as memories age) results in the loss of spatio-temporal information. Empirical support for the notion that spatial specificity is lost as memories age includes rodent studies by Wiltgen et al. (2010), Gafford, Parsons, & Helmstetter (2013) and Einarsson et al. (2015). On the other hand, Dede and Smith, in a comprehensive review of human cases of temporal lobe damage (Dede & Smith, 2016), conclude that there is no evidence for loss of remote episodic information after hippocampal damage, contradicting the transformation hypothesis and supporting the standard theory.

While the standard theory thus is not universally accepted (Nadel & Moscovitch, 1997; Rudy, Biedenkapp, & O'Reilly, 2005; Winocur, Moscovitch, & Bontempi, 2010), it enjoys widespread support and is compatible with a large body of empirical evidence (Dede & Smith, 2016; Runyan, Moore, & Dash, 2019; Squire, Genzel, Wixted, & Morris, 2015; Wang & Morris, 2010; Wiltgen, Brown, Talton, & Silva, 2004).

Several neocortical regions in the frontal and temporal lobes have been identified as locations where memories consolidate, including the frontal cortex, posterior and anterior cingulate and temporal cortex (Bontempi, Laurent-Demir, Destrade, & Jaffard, 1999; Frankland, Bontempi, Talton, Kaczmarek, & Silva, 2004; Quirk & Mueller, 2008), parietal and retrosplenial cortex (Maviel, Durkin, Menzaghi, & Bontempi, 2004) (both using a spatial discrimination task in rodents), and frontal, lateral temporal, and occipital lobes for autobiographical memories in humans (Bayley, Gold, Hopkins, & Squire, 2005). For reviews, see Dede and Smith (2016) and Frankland and Bontempi (2005). The anterior cingulate cortex (ACC), part of the prefrontal cortex, has received particular attention because it has been shown to play an important role for recall of remote memories in several experimental paradigms including fear conditioning, conditioned taste aversion, trace eyeblink conditioning and spatial discrimination (Bontempi et al., 1999; Ding, Teixeira, & Frankland, 2008; Einarsson & Nader, 2012; Einarsson et al., 2015;

Frankland & Bontempi, 2005; Frankland et al., 2004; Maviel et al., 2004; Takehara, Kawahara, & Kirino, 2003).

**Experimental methods.** Much information regarding the role of the hippocampus in systems consolidation comes from lesion experiments with rats or mice, where bilateral hippocampal lesions are performed at different intervals following training. In a common type of fear conditioning, a sound or a specific spatial context (conditioned stimulus) is paired with an electric foot shock or other aversive unconditioned stimulus. Recall is subsequently tested by presenting the conditioned stimulus alone and measuring the degree of fear response. Findings from this type of study indicate that hippocampal lesions immediately after training result in severely impaired recall, but longer delays between conditioning and lesions produce gradually less severe impairments. The time interval after conditioning during which hippocampal lesions result in significant memory impairment is called the *systems consolidation window*. For fear conditioning in rodents, most studies report a consolidation window of between three and four weeks (J. Kim, Clark, & Thompson, 1995; J. Kim & Fanselow, 1992; Kitamura et al., 2009; Land, Bunsey, & Riccio, 2000; Winocur, Moscovitch, & Sekeres, 2013). Longer windows have been reported for primates: months in monkeys (Zola-Morgan & Squire, 1990), and years in humans (Scoville & Milner, 1957; Squire & Cohen, 1979).

Systems consolidation has also been investigated by studying the effect of reversible inactivation of specific brain areas. Studies using pharmaceutical inactivation of the rodent hippocampus and/or ACC have shown that retrieval of a fear memory is hippocampus-dependent one or three days after acquisition, but not after 28 or 30 days. At this point it has instead become dependent on the ACC for retrieval (Einarsson et al., 2015; Frankland et al., 2004; Sierra et al., 2017; Wiltgen et al., 2010).

Thus results from both lesion and inactivation studies are compatible with the standard theory, and similar results have also been obtained in studies measuring brain activity during retrieval of recent and remote memories (Bontempi et al., 1999; Gafford et al., 2013; Nadel & Hardt, 2010).

**Reconsolidation.** Whereas synaptic reconsolidation – the transient post-reactivation susceptibility of memories to amnestic interventions like electro-convulsive shock or administration of protein synthesis inhibitors – has been studied since the 1960s (Judge & Quartermain, 1982; Lewis et al., 1972; Misanin et al., 1968), systems reconsolidation – the return of hippocampus-dependence after reactivation – was first described by Land et al. in 2000 (Land et al., 2000), and subsequently by Debiec et al. (2002) and Winocur et al. (2009; 2013). These researchers found that hippocampal lesions produced amnesia for 30- or 45-day old fear memories if performed immediately after reactivating the memories by presentation of the conditioned stimulus. Hippocampal lesions did not produce memory impairments without preceding reactivation, nor if administered after the reactivated memory was allowed 48 hours to restabilize after reactivation (Debiec et al., 2002). It thus appears that reactivation renders a neocortical memory trace unstable and that a functioning hippocampus is needed for its restabilization.

Inactivation studies have provided additional information about the effect that reactivation has on remote memories. As noted above, 30-day old fear memories are strongly dependent on ACC for retrieval. In an elegant study, Einarsson et al. (2015) showed that six hours after memory reactivation, ACC inactivation no longer impaired retrieval, but after 24 hours ACC-dependence had returned. At the six-hour time point, inactivation of the hippocampus also did not affect retrieval, but simultaneous activation of both hippocampus and ACC did block recall.

Together, the findings from post-reactivation lesion and inactivation studies suggest that reactivation triggers the creation of a short-lived hippocampal memory trace that is able to support recall and is also required for restabilization of the ACC trace.

An additional interesting finding is that PSI infusion into hippocampus immediately after reactivation blocks reconsolidation when performed three days (Debiec et al., 2002; 2006), five days (Rossato, Bevilaqua, Medina, Izquierdo, & Cammarota, 2006), or seven days (Haubrich et al., 2015; Milekic & Alberini, 2002) after training. These results suggest that recovery from the destabilization caused by reactivation depends on protein synthesis in the hippocampus.

Researchers differ regarding the maximum memory age at which reactivation can trigger systems reconsolidation. Whereas the lesion and inactivation studies indicate that memories are susceptible for at least a month after training (Debiec et al., 2002; Einarsson et al., 2015; Land et al., 2000; Winocur et al., 2009; Winocur, Sekeres, et al., 2013), the results from post-reactivation PSI infusion into the hippocampus are less clear. Debiec et al. (Debiec et al., 2002) were able to demonstrate reconsolidation blockade with anisomycin as late as 45 days after training, whereas Frankland et al. (Frankland et al., 2006) were unable to show the effect at 36 days. The reason for this discrepancy is not clear.

An interesting clue to a connection between synaptic and systems reconsolidation is provided by Ghazal (2016): reactivation of a 1-day-old fear memory triggered a strong reduction of CI-AMPARs in hippocampus but not in ACC, whereas at 30 days the opposite was true. This result is consistent with a transition from engagement of the hippocampus to the ACC over a systems-consolidation timeframe and with synaptic reconsolidation taking place in whichever of the two systems is engaged in the retrieval.

In summary, retrieval of a new (e.g. 3-day-old) fear memory requires the hippocampus but not the ACC. Over time, a reversal takes place so that retrieval of a 30-day-old memory requires the ACC but not the hippocampus. Reactivation of a consolidated memory temporarily returns it to ACC-independence for retrieval. Systems consolidation (establishment of a neocortical trace) and systems reconsolidation (restabilization after reactivation-induced destabilization of the neocortical trace) both require hippocampal involvement.

## Model

Based on the findings described in the foregoing, we here present a model of systems memory consolidation and reconsolidation. Although the role of LTP in learning and memory remains conjectural, for the purpose of modeling we assume that LTP – or an LTP-like mechanism – is the cellular substrate of memory and investigate to what extent this type of synaptic mechanism can explain the systems-level phenomena.

Below we provide a conceptual description of the model; the computational implementation is described in the Methods section.

### Synaptic level

- Moderately intense stimulation induces E-LTP, which involves the rapid insertion of CP-AMPARs. Constitutive processes subsequently remove these within hours.
- More intense stimulation sets in motion L-LTP induction (synaptic consolidation) which involves a state change in a bistable mechanism (molecular switch). When in the ON state, this mechanism maintains a high CI-AMPAR count in the synapse;

when the switch is OFF, the CI-AMPAR count drifts towards the basal level characteristic of the unpotentiated synapse.

- Memory retrieval abruptly removes CI-AMPARs from the synapse and replaces them with CP-AMPARs, thus returning the synapse to an E-LTP-like state. The subsequent restoration of L-LTP is protein-synthesis-dependent and requires NMDAR activity, i.e. neural stimulation.

**Systems level**

As discussed above, some researchers disagree with the standard theory, in particular with respect to episodic/autobiographical memories. Nevertheless, there is widespread agreement that semantic information does undergo systems consolidation. The aim of the present work is to model those types of memories that do undergo a reorganization whereby they gradually become hippocampus-independent; the question of whether episodic/autobiographical memories are processed differently is beyond our present scope.

- Stimulus presentations trigger patterns of activation in multiple ensembles of neurons in the neocortex (NC). These active neurons in turn project onto and activate neurons in the hippocampus (HPC), where a memory trace is quickly created, providing linkages between the activated NC ensembles. Linkages are also created through prefrontal cortex, in particular the ACC, but these connections are initially too weak to support memory retrieval without HPC support.
- Subsequently, the HPC memory trace is spontaneously and repeatedly reactivated which causes stimulation of these same NC neural ensembles through nerve fibers projecting back from the HPC to the NC. Over time, the repeated reactivation of the NC neural ensembles strengthens the ACC linkages, eventually to a point where they can support retrieval of the memory without assistance from the hippocampus.

- Meanwhile, the HPC trace is gradually weakened by constitutive decay processes (Hardt, Nader, & Nadel, 2013; Sachser et al., 2016).
- If a consolidated memory is reactivated by a reminder, then activity in the neocortical neural ensembles triggers re-establishment of the HPC linkage. Simultaneously, the retrieval causes transient destabilization of the synapses in the ACC linkage (synaptic reconsolidation).
- Following reactivation, hippocampal replay stimulates the now destabilized synapses of the ACC linkage. This activity drives restabilization of these synapses. Meanwhile, the reactivated HPC trace rapidly decays, leading to a return to ACC-dependence in 24 hours or less.

## Temporal characteristics

The temporal characteristics of systems consolidation and reconsolidation provide some clues to the underlying mechanisms:

***The reconsolidation window is shorter than the consolidation window.*** Hippocampal lesion after training produces a memory impairment, more severe the shorter the delay between conditioning and lesion. Similarly, hippocampal lesion after reactivation also causes more severe impairment if performed early. However, the time scales are very different. New memories are vulnerable for at least several weeks after training (J. Kim & Fanselow, 1992; Squire & Cohen, 1979; Winocur, 1990; Zola-Morgan & Squire, 1990), whereas the post-reactivation window of vulnerability only lasts for a few days (Debiec et al., 2002). Our model attributes this difference to the different natures of the processes being interrupted by HPC lesioning in the two scenarios. In the case of consolidation, what is being interrupted is the gradual and relatively slow establishment and strengthening of intra-neocortical connections; thus the earlier the intervention is performed, the weaker the partially consolidated memory trace will be. Reconsolidation blockade, on the other hand, interrupts the much faster process of re-stabilization of destabilized neocortical synapses. The earlier in the reconsolidation window hippocampal lesion is performed, the fewer synapses will have had time to restabilize, leading to more severe memory loss.

***Memories become transiently ACC-independent after reactivation.***

Whereas a consolidated fear memory depends strongly on the ACC for recall, reactivation triggers a brief period of ACC-independence, such that six hours after reactivation ACC inactivation has little or no effect on retrieval, but 24 hours after reactivation full ACC dependence has returned (Einarsson et al., 2015). This finding suggests that reactivation triggers the creation of a short-lived hippocampal memory trace that is able to support recall six hours later, but not 24 hours later. For a memory trace to last for six hours or more, it must have undergone synaptic consolidation, yet 24 hours is a short lifetime for a consolidated trace – compare the situation after initial acquisition, where the hippocampal trace is able to support recall for at least several days. The mechanism that causes a faster decline of a post-reactivation hippocampal trace is unknown.

**Previously Published Computational Models**

Several computational models of systems consolidation have been published. Alvarez and Squire (1994) described a recurrent artificial neural network (ANN) model with three memory banks representing, respectively, the medial temporal lobe (MTL - the hippocampus and adjacent areas) and two areas of neocortex (NC). Learning was implemented by a Hebbian-like algorithm that adjusted connection weights, with a fast learning rate for the connections between the MTL and each of the NC areas (MTL-NC connections) and a slower rate for the connections between the two NC areas (NC-NC connections). After training on a pair of random patterns, one in each NC area, the model could be presented with either of the two patterns and reproduce the other with good accuracy thanks to the strong linkage provided by the MTL-NC connections, but not if the MTL subsystem was disabled. However, if after initial training, MTL units were repeatedly reactivated in random fashion, then the resulting stimulation through the strong MTL-NC connections would cause the two trained NC patterns to become active, and Hebbian learning would strengthen the NC-NC connections to a point where they eventually became able to support recall without the MTL linkage. The Alvarez and Squire model thus demonstrated the computational feasibility of the standard theory of systems consolidation.

McClelland et al. (1995) hypothesized that systems consolidation serves to avoid “catastrophic interference”, impairment of older memory traces when new patterns are integrated too quickly, an effect that had been observed in artificial neural networks. They suggested that the brain avoids this problem by having the hippocampus quickly capture new patterns, then repeatedly replaying them to the slow-learning neocortex. A computational model described in the paper demonstrated that such repeated replay, interleaved with rehearsal of previously learned patterns, could avoid catastrophic interference. The model used in their simulations consisted of a traditional three-layer feed-forward network.

Meeter and Murre's *TraceLink* model (Meeter & Murre, 2005; Murre, 1996) combines elements from these two earlier studies. Like Alvarez and Squire's (1994) model, TraceLink is a recurrent network with neocortical and hippocampal layers. Unlike the Alvarez and Squire model, it is capable of learning multiple patterns in succession, avoiding catastrophic interference by repeated activation of previously learned material. It differs from the McClelland et al. (1995) model by including simulation of a hippocampal system which autonomously presents a random mix of new and old patterns to neocortex. McClelland et al.'s simulation did not include a hippocampus component; the interleaved set of old and new training patterns was provided to the network by the programmer.

A paper by Amaral et al. (2008) is primarily concerned with demonstrating that a single model can accommodate seemingly contradictory results with respect to the transience or permanence of amnesia resulting from hippocampal lesions at different time points in the course of systems consolidation. The model consists of two recurrent layers, representing hippocampus and neocortex and employs the "synaptic reentry reinforcement" (SRR) mechanism, first described by Wittenberg and Tsien (2002), which can autonomously strengthen a memory trace over time, and even repair partial damage. The model is able to reproduce the empirical finding that both early and late hippocampal disruption can produce either permanent or transient memory impairment, but permanent loss is a more likely outcome after early and/or extensive hippocampal disruption.

These simulations thus illuminate various aspects of systems consolidation. Our work builds on these previous models and adds the capability to simulate systems reconsolidation, i.e.

a transient period of hippocampus-dependence following cued retrieval of a consolidated memory. The key to this capability is a more elaborate connection design than is traditionally used in neural networks, Specifically, the connections in our model simulate the AMPA receptor exchanges underlying synaptic consolidation and reconsolidation. Below, we introduce our computational implementation of the model described above, and show that it is capable of reproducing both systems consolidation and reconsolidation, including the effects of different pharmaceutical and surgical interventions.

## Methods

### Simulation Targets

Table 1 summarizes the empirical findings, as described in the introduction, which our model aims to reproduce in simulations.

Table 1: Simulation targets

| | Result | Description |
|---|---|---|
| **1** | **Retrieval is HPC-dependent and ACC-independent at 3d.[1,2,3,4,5]** | **Retrieval of a three-day old memory is impaired by HPC inactivation but unaffected by ACC inactivation.** |
| **2** | Retrieval is ACC-dependent and HPC-independent at 30d.[1,2,3,4,6] | Retrieval of a 30-day old memory is impaired by ACC inactivation but unaffected by HPC inactivation. |
| **3** | PSI during conditioning impairs LTM but not STM.[5,7,8] | Systemic injection of PSI during training prevents LTM induction – but does not impair STM . |
| **4** | PSI infusion during maintenance does not cause memory impairment.[5,7] | Systemic PSI injection does not impair a consolidated memory. |
| **5** | HPC lesion at 3d causes memory impairment.[9,10] | HPC lesion at 3 days or less after training causes later recall to be severely impaired compared to non-lesioned animals. |
| **6** | HPC lesion at 30d does not cause memory impairment.9,10,11 | HPC lesion, when not preceded by memory reactivation, does not result in subsequently impaired recall. |
| **7** | Reactivation alone does not impair a consolidated memory.[10,11] | Reactivation without subsequent HPC lesion or PSI infusion does not by itself impair a consolidated memory. |
| **8** | Reactivation + PSI causes memory impairment.[8,11,12,13,14] | Reactivation immediately followed by PSI infusion into HPC causes impairment of a consolidated memory. |
| **9** | Delayed impairment effect of post-reactivation PSI infusion in HPC.[8,11] | The impairment caused by post-reactivation PSI infusion in HPC does not manifest in retrieval test soon after lesion (4h) but only later (24h). |
| **10** | Reactivation + HPC lesion causes memory impairment.[10,11,15] | Reactivation immediately followed by HPC lesion causes impairment of a consolidated memory. |
| **11** | Graded effect of post-reactivation HPC lesion.[11] | The severity of memory impairment caused by post-reactivation HPC lesion diminishes with increasing reactivation-lesion delay. |
| **12** | Retrieval can be supported by either ACC or HPC 6h after reactivation.[1] | Six hours after reactivation, neither ACC inactivation nor HPC inactivation alone impairs retrieval, but simultaneous inactivation of both does block retrieval. |

[1]Einarsson et al. (2015) [2]Gafford et al. (2013) [3]Ghazal (2016) [4]de Oliveira Alvares et al. (2012) [5]Frey (1997) [6]Frankland et al. (2004) [7]Davis and Squire (1984) [8]Rossato et al. (2006) [9]J. Kim and Fanselow (1992) [10]Land et al. (2000) [11]Debiec et al. (2002) [12]Frankland et al. (2006) [13]Haubrich et al. (2015) [14]Milekic and Alberini (2002) [15]Winocur et al. (2009)

## Model Design

Like other artificial neural networks, ours consists of units and connections, where units are analogs of biological neurons (or ensembles of neurons) and connections model synapses. (For a primer on artificial neural networks, see e.g. Shultz (2003)).

**Topology.** We use a recurrent artificial neural network with four *regions* representing hippocampus (HPC), anterior cingulate cortex (ACC) and two sensory cortex areas, SC0 and SC1, to which stimuli are presented. Each region consists of 25 *units*. The units are linked together by *connections*, such that each HPC and ACC unit is bidirectionally connected to all the units in the other three regions. The set of connections between any two regions, e.g. from HPC to SC1, is referred to as a *tract*, see Figure 1.

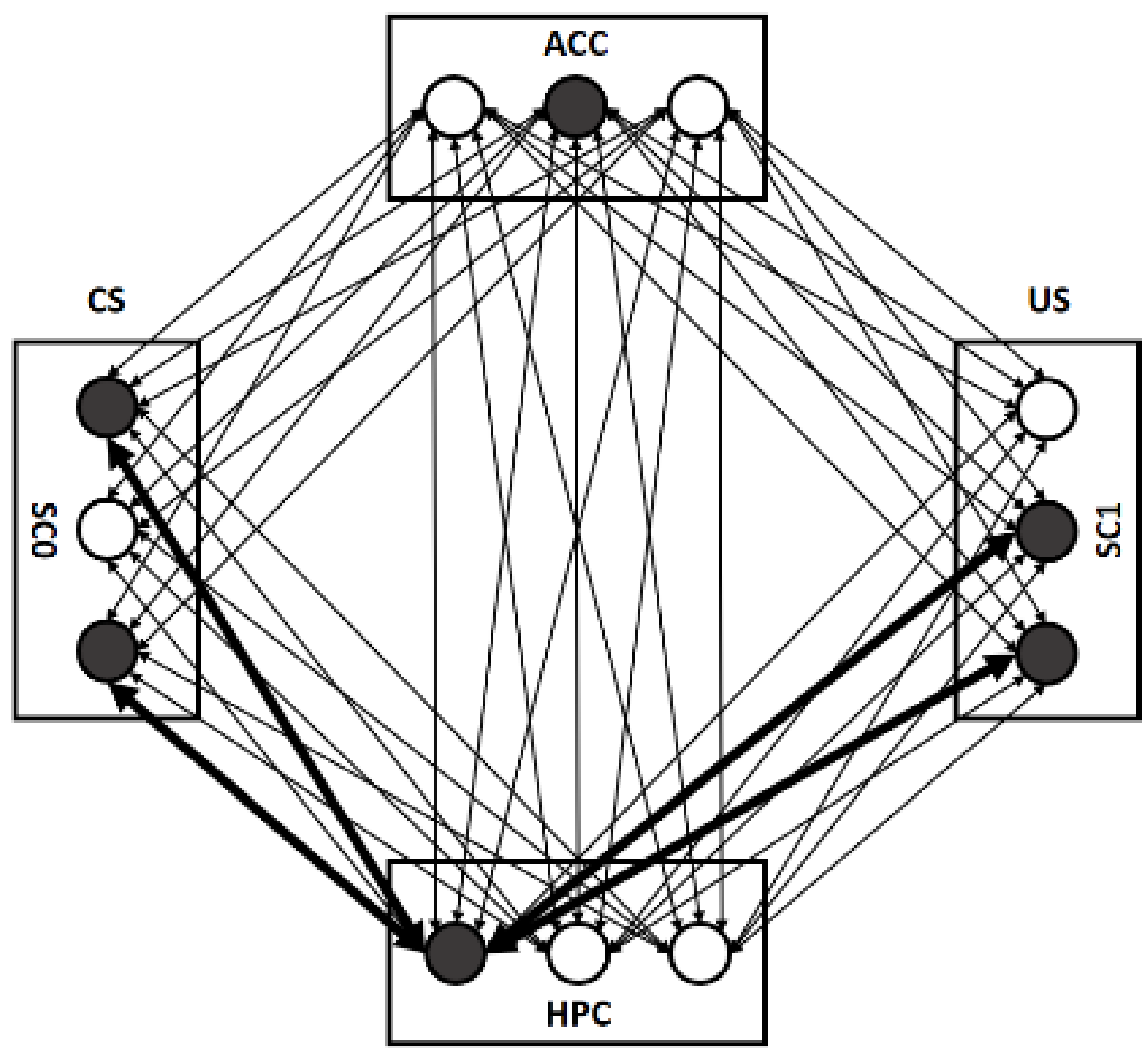

Figure 1: Network architecture. To reduce clutter only three of the twenty-five units in each region are shown. Each double-headed arrow represents two independent connections, one in each direction, between a pair of units. The diagram illustrates the state after initial acquisition: presentation of the unconditioned stimulus (US) and the conditioned stimulus (CS) has activated some units in SC0 and SC1 (filled circles) and fast learning has created strong linkages (bold lines) through the HPC. Linkages through the slower-learning ACC are still weak.

This topology is an evolution of the one used by Alvarez and Squire in their influential model of systems memory consolidation (Alvarez & Squire, 1994). Their model had two neocortical regions with direct linkage between them; we have added the ACC region in order to be able to simulate interruption of linkage between the SC1 and SC2 regions by reversible inactivation of either the ACC and/or HPC regions (or both).

As noted in the introduction, a number of neocortical regions other than the ACC have been identified as loci for remote memory traces. For simplicity, the model only includes the ACC as a target region for systems consolidation.

Similarly, the model's connectivity between the hippocampus and neocortical regions is simplified for reasons of computational feasibility. In a real brain, the hippocampal complex is comprised of many intricately interconnected subareas that receive input from and project back to not only sensory cortices, but to a widespread network of different structures in hierarchies of association areas. For an overview, see Lavenex and Amaral (2000).

The algorithms for activation, learning, and retrieval are adapted from Meeter and Murre's *TraceLink* model of systems memory consolidation (Meeter & Murre, 2005), with minor modifications that will be detailed in the following.

The model description that follows makes reference to a number of configurable parameters. These are documented in tables 2-4.

**Units.** The units are bistable and stochastic. A unit's activity level is either 1 (active) or 0 (inactive). The probability that a unit is active (the *activation function*) is an asymmetric sigmoid function of its net input:

$$P(isActive_j) = \frac{1}{1+e^{-actK(net_j - inhib_r)}} \quad (1)$$

Where $P_j(isActive_j)$ is the probability that unit *j* is active, and $net_j$ is the net input to unit *j*, calculated as the sum of the activity levels of units connected to unit *j*, weighted by inbound connection strengths:

$$net_j = \sum_i a_i s_{ij} \quad (2)$$

Here, $a_i$ is the activation level of unit *i* and $s_{ij}$ is the strength of the connection from unit *i* to unit *j*. The *actK* symbol in Equation 1 is a configurable parameter that controls the slope of the sigmoid function. The $inhib_r$ variable in Equation 1 is a region-specific dynamic inhibition level, which will be described below.

Figure 2 shows two examples of activation functions with *actK* values of 2.0 and 10.0, respectively, both with an inhibition level of 5.0. The probability of activation grows from near zero when net input is low to near one when net input is high, with a value of 0.5 at the current inhibition level. An increase in inhibition shifts the curve to the right, with the effect that greater net input is required for activation. This tends to results in fewer active units in the region. Greater *actK* values produce curves with steeper, more abrupt transitions, corresponding to a more deterministic behavior. All simulations reported here use *actK* = 2.0.

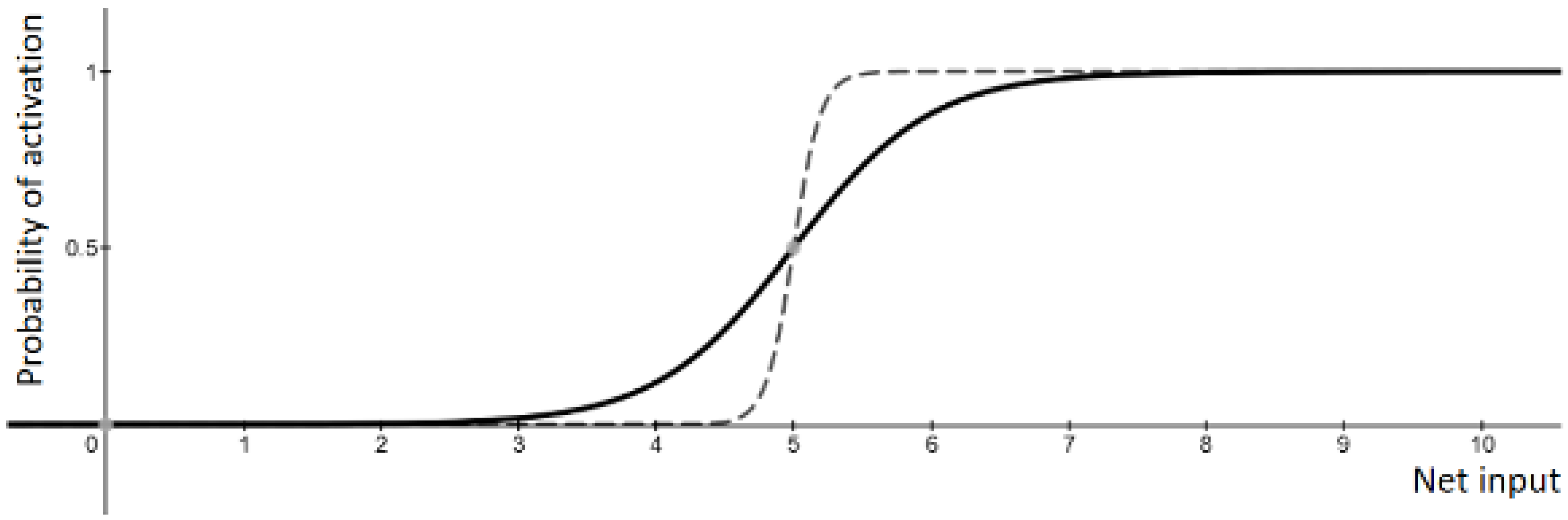

Figure 2: Activation functions. Two examples are shown, one with *actK = 2.0* (solid) and one with *actK = 10.0* (dashed). Both are plotted with $inhib_r = 5.0$.

The use of the activation function, including the inhibition mechanism, will be further described below.

**Connections.** The connections are abstract models of glutamatergic synapses, characterized by four variables:

- *psdSize*: the maximum number of AMPARs that can be inserted, analogous to the number of receptor "slots" in the PSD.
- *numCpAmpars*: the number of currently inserted CP-AMPARs
- *numCiAmpars*: the number of currently inserted CI-AMPARs
- *isPotentiated*: a Boolean attribute that models the bistable nature of L-LTP

This combination of binary and continuously variable attributes makes it possible to model synapses where potentiation behaves like a bistable switch (Bortolotto, Bashir, Davies, & Collingridge, 1994; Jalil et al., 2015; Westmark et al., 2010), yet synaptic strength is variable with many distinct levels (Bartol et al., 2015).

As will be described below, stimulation causes an increase of the *psdSize* attribute, allowing more AMPARs to be inserted. A connection's weight at any moment is proportional to its total number of inserted AMPARs (*numCpAmpars* + *numCiAmpars*).

## Simulation

A simulation consists of a sequence of time steps. Various interventions may be scheduled for any time point during the simulation, and in addition several background processes execute at each time step. The scheduled event types are training, reactivation, HPC lesion, HPC inactivation and ACC inactivation. The background processes are consolidation, AMPAR trafficking and random depotentiation. In addition, a retrieval test can be executed at any time. The different interventions and background processes will be described below, but first we introduce the concept of a learning cycle.

**Learning cycle.** Learning takes place when stimuli are presented, when a memory is reactivated, and during memory consolidation. In each of these instances, learning is achieved by the execution of a *learning cycle* in every connection that satisfies the Hebbian condition that the two units that it connects are both active. A learning cycle consists of three steps: PSD growth, CP-AMPAR influx, and (possibly) L-LTP induction.

**1. PSD growth.** The network learns activation patterns by a Hebbian learning rule (Hebb, 1949) that increases the *psdSize* attribute of connections between simultaneously activated units, asymptotically towards a maximum value:

$$psdSize_{ij} \leftarrow psdSize_{ij} + \mu_T(maxPsdSize - psdSize_{ij}) \tag{3}$$

where $psdSize_{ij}$ is the PSD size of the connection between units *i* and *j*, $maxPsdSize$ is a configurable parameter, and $\mu_T$ is a learning rate specific to the tract that connection *ij* belongs to. This learning rule differs from that of TraceLink in three respects: (1) We use an asymptotic function where TraceLink used a linear function with hard limits; this is an esthetic choice; (2) In TraceLink, the learning rule applies directly to connection weights; in our model, learning affects PSD size, i.e. the maximum weight that the connection can attain; actual weight is defined by the number of inserted AMPARs, which may at times be lower than psdSize; (3) TraceLink's learning rule supported "unlearning" (learning at a negative rate) to model interference, a phenomenon that we are not addressing in the present model.

To simulate different intensities of stimulation, the learning rule is applied a variable number of times, specified by the parameter *numStimCycles*. The stimulation resulting from spontaneous hippocampal activation during consolidation is modeled as much less intense than the stimulation due to stimulus presentation.

**2. CP-AMPAR influx.** PSD growth is followed by an increase in the number of CP-AMPARs such that the total AMPAR count becomes equal to the connection's PSD size:

$$numCpAmpars_{ij} \leftarrow psdSize_{ij} - numCiAmpars_{ij} \tag{4}$$

This models the rapid CP-AMPAR influx during E-LTP induction.

**3. L-LTP induction.** Finally, probabilistic induction of L-LTP in a connection is implemented by turning on its *isPotentiated* attribute with a probability that depends on the intensity of the stimulation:

$$prob \leftarrow f(numStimCycles, stimThresh) \tag{5}$$

where *numStimCycles* is as described above, *stimThresh* defines the value of *numStimCycles* that produces a 0.5 probability of L-LTP induction, and *f* is an asymmetric sigmoid function:

$$f(stim, thresh) = \frac{1}{1 + e^{-(stim - thresh)}} \tag{6}$$

Learning cycles are executed (a) when stimuli are presented for training, (b) at memory retrieval (reactivation), and (c) when patterns are spontaneously activated by the memory consolidation process. Descriptions of these mechanisms follow:

**Training.** To train an association, patterns (subsets) of units are activated in the SC0 and SC1 regions to represent an unconditioned stimulus, US, and a conditioned stimulus, CS, respectively. Each stimulus pattern consists of $k_r * numUnits_r$ units, where $numUnits_r$ is the number of units in region *r* and $k_r$ is a region-specific parameter that specifies the fraction of the region's units to activate in a pattern. (This is described further in the section about inhibition, below.) The network then randomly selects and activates $k_r$ linkage units in each of the HPC and ACC regions and executes a learning cycle in all connections that connect two simultaneously active units, i.e. connections that are in the Hebbian condition. The learning rate, $\mu_T$, is defined to be relatively high in HPC tracts, allowing rapid creation of linkages strong enough to support recall. The ACC learning rate is lower, hence linkages through the ACC region are not strong enough to independently support recall immediately after training.

**Retrieval.** To test recall of a trained pattern, the CS units are activated in SC0, and the network is cycled by repeated application of the activation function in all units a number of times specified by the parameter *numSettleCycles*. The units are updated synchronously: new activation values are first computed for all units, based on their current net inputs. The new values are then applied to all the units.

The activity pattern that the SC1 region eventually settles into is compared to the associated US pattern to calculate a retrieval test score.

**Inhibition.** To ensure fast settling of the network, we use a variation of an inhibition mechanism described by Meeter and Murre for their *TraceLink* architecture, designed to (roughly) model the working of inhibitory neurons (Meeter & Murre, 2005). Each region has a configurable target fraction of active units, $k_r$. Following each update cycle during the settling phase of memory retrieval (and during consolidation, see below), each region's inhibition level, $inhib_r$ (see Equation 1), is adjusted by an amount proportional to the current deviation of the number of active units in the region from the target value defined by the region's $k_r$ value:

$$inhib_r \leftarrow inhib_r + \frac{numActive_r - k_r * numUnits_r}{k_r} * inhibIncr \quad (7)$$

$$inhib_r \leftarrow min(maxInhib(max(minInhib, inhib_r))) \quad (8)$$

where $numActive_r$ is the number of active units in region *r* and $numUnits_r$ is the total number of units in region *r*. The configurable *inhibIncr* parameter controls the rate of adjustment of the inhibition level. The assignment in Equation 8 restricts $inhib_r$ to a configurable range of values $[minInhib, maxInhib]$.

**Systems consolidation.** At every simulation time step a randomly selected trained pattern is activated in HPC, after which the entire network is cycled in the same manner as for recall test (but without stimulus presentation). Whatever pattern the ACC and SC regions settle into is then reinforced by application of the learning rule. Because the network is more likely to settle into trained patterns than other random states, this will tend to strengthen CS-US linkages through the ACC, eventually making recall of trained patterns HPC-independent.

**AMPAR trafficking.** At each simulation time step, AMPAR trafficking is simulated in all connections as follows:

CP-AMPARs: the *numCpAmpar* value declines asymptotically towards zero, simulating the limited dwell time of CP-AMPARs at the synapse:

$$numCpAmpars_{ij} \leftarrow numCpAmpars_{ij} - cpAmparRemovalRate * (numCpAmpars_{ij} - minNumCpAmpars) \quad (8)$$

where $numCpAmpars_{ij}$ is the current number of CP-AMPARs at connection *ij*, and *cpAmparRemovalRate* and *minNumCpAmpars* are configurable parameters.

CI-AMPARs: if a connection's *isPotentiated* attribute is true and its source and destination units are both active (Hebbian condition), then *numCiAmpars* grows asymptotically towards the number of available slots in the connection (*psdSize*), simulating activity-dependent trafficking of CI-AMPARs into a potentiated synapse:

$$numCiAmpars_{ij} \leftarrow min(numCiAmpars_{ij} + ciAmparInsertionRate, psdSize_{ij}) \quad (9)$$

In unpotentiated connections, $numCiAmpars_{ij}$ declines exponentially, simulating a depotentiated synapse's gradual return to baseline:

$$numCiAmpars_{ij} \leftarrow numCiAmpars_{ij} - ciAmparRemovalRate * (numCiAmpars - minNumCiAmpars) \quad (10)$$

*minNumCiAmpars, ciAmparInsertionRate* and *ciAmparRemovalRate* are configurable parameters.

**Depotentiation.** Potentiated connections are subject to random constitutive depotentiation at a rate controlled by the tract-specific parameter $baseDepotProb_t$ ,which has a higher value for HPC tracts than for intra-neocortical tracts, to account for the observed faster decline of hippocampal memory traces.

**Reactivation.** Reactivation is modeled as an unreinforced CS presentation, i.e. a cued retrieval. The CS pattern is activated in the SC0 region, the network is settled as described for training, and an AMPAR exchange is simulated in all connections situated between simultaneously active units (i.e. connections in the Hebbian condition): $numCiAmpars_{ij}$ is reduced to its minimum value *minNumCiAmpars*, and $numCpAmpars_{ij}$ is set to its maximum allowed value $psdSize_{ij} - numCiAmpars_{ij}$. This puts those consolidated ACC linkage connections that were activated by memory retrieval in an unstable E-LTP-like state, modeling the post-reactivation instability documented in empirical studies. A random set of $k_r$ HPC linkage units is then activated, in the same manner as during initial acquisition, and a round of Hebbian learning takes place.

As noted in the introduction, the hippocampal engagement after reactivation is much briefer (less than 24h) than after initial training, when HPC can support recall for several days. Although the mechanism underlying this faster disengagement is unknown, in our model we simulate the phenomenon by increasing the depotentiation probability in connections that are activated following memory retrieval:

$$depotProb_{ij} \leftarrow maxDepotProb_t \qquad (11)$$

where $maxDepotProb_t$ is a tract-specific parameter that has a higher value in HPC tracts than in neocortical tract. Subsequently, the depotentiation rate returns asymptotically to its base value:

$$depotProb \leftarrow depotProb - depotProbDecayRate * (depotProb - baseDepotProb) \qquad (12)$$

where *depotProbDecayRate* is a global parameter.

**Hippocampal lesion.** Hippocampal lesion is simulated by disconnecting the HPC region from the simulation.

**PSI infusion.** Infusion of PSI into a region – HPC or ACC – is simulated by disabling potentiation and CI-AMPAR insertion for nine hours of simulation time, corresponding to the amount of time that the protein synthesis inhibitor anisomycin remains active in brain tissue (Wanisch & Wotjak, 2008).

**Inactivation.** Reversible inactivation of HPC or ACC is modeled by transiently disabling activation of all units in the HPC or ACC region, respectively.

**Simulation environment**

The model is implemented as a C++ program and all simulations were executed on an Intel i5-2400 computer running the Debian Linux 8.4 operating system.

**Model fitting**

Simulations are controlled by the parameters indicated in the following tables. Parameter values were manually chosen to make simulation time courses approximate those reported in the referenced literature.

Table 2: Global parameters

| **Parameter** | **Value** | **Description** |
|---|---|---|
| *actK* | 2.0 | Steepness of the activation function |
| *minPsdSize* | 10 | Min value for a connection's psdSize |
| *maxPsdSize* | 100 | Max value for a connection's psdSize |
| *numSettleCycles* | 20 | Number of times the network is cycled for settling |
| *trainNumStimCycles* | 50 | Number of times the learning rule is applied at training |
| *consNumStimCycles* | 1 | Number of times the learning rule is applied at consolidation |
| *minInhib* | 2.5 | Minimum inhibition level in a region |
| *maxInhib* | 10 | Maximum inhibition level in a region |
| *inhibIncr* | 0.05 | Inhibition level adjustment rate |

Table 3: Region-specific parameters.

| **Parameter** | **HPC** | **ACC** | **SC0/1** | **Description** |
|---|---|---|---|---|
| $numUnits_r$ | 25 | 25 | 25 | Number of units |
| $k_r$ | 0.2 | 0.2 | 0.2 | Target fraction of active units |

Table 4: Tract-specific parameters. The “HPC” columns indicates values for the tracts that connect HPC with the other regions, and the “ACC” column for the tracts that connect ACC with SC0 and SC1.

| **Name** | **HPC** | **ACC** | **Description** |
|---|---|---|---|
| $\mu_t$ | 0.08 | 0.004 | Learning rate |
| *psdDecayRate01h* | 0.01 | 0.01 | PSD shrink rate when unpopulated |
| *cpAmparRemovalRate01h* | 0.1 | 0.1 | Constitutive removal rate for CP-AMPARs |
| *ciAmparInsertionRate01h* | 2.0 | 2.0 | CI-AMPAR insertion rate in potentiated synapse |
| *ciAmparRemovalRate01h* | 0.015 | 0.015 | CI-AMPAR removal rate in unpotentiated synapse |
| *baseDepotProb01h* | 0.002 | 0.0 | Probability of spontaneous depotentiation |
| *maxDepotProb01h* | 0.05 | 0.0 | Prob. of depotentiation after reactivation |
| *depotProbDecayRate01h* | 0.03 | 0.03 | depotProb decay rate towards baseDepotProb |
| *minNumCpAmpars* | 0 | 0 | Num CP-AMPARs in an unpotentiated conn. |
| *minNumCiAmpars* | 2 | 2 | Num CI-AMPARs in an unpotentiated conn. |

## Results

After training the network with a CS-US association, recall is tested by presenting the CS in the SC0 region, settling the system, and comparing the resulting activation pattern in the SC1 region with the trained US (a score of 1.0 indicates a perfect match). The descriptions below compare the recall scores in simulation runs where an intervention (lesion or inactivation) is performed with simulation runs without the intervention (labeled “baseline” in the diagrams). The score values in all diagrams are means of 100 simulation runs. Error bars indicate standard deviation.

**Consolidation and reconsolidation windows.** HPC lesions produce memory deficits when performed in the consolidation or reconsolidation windows, but not otherwise. See Figure 3 and Figure 4. These simulations reproduce findings 5-7 and 10-11 of Table 1.

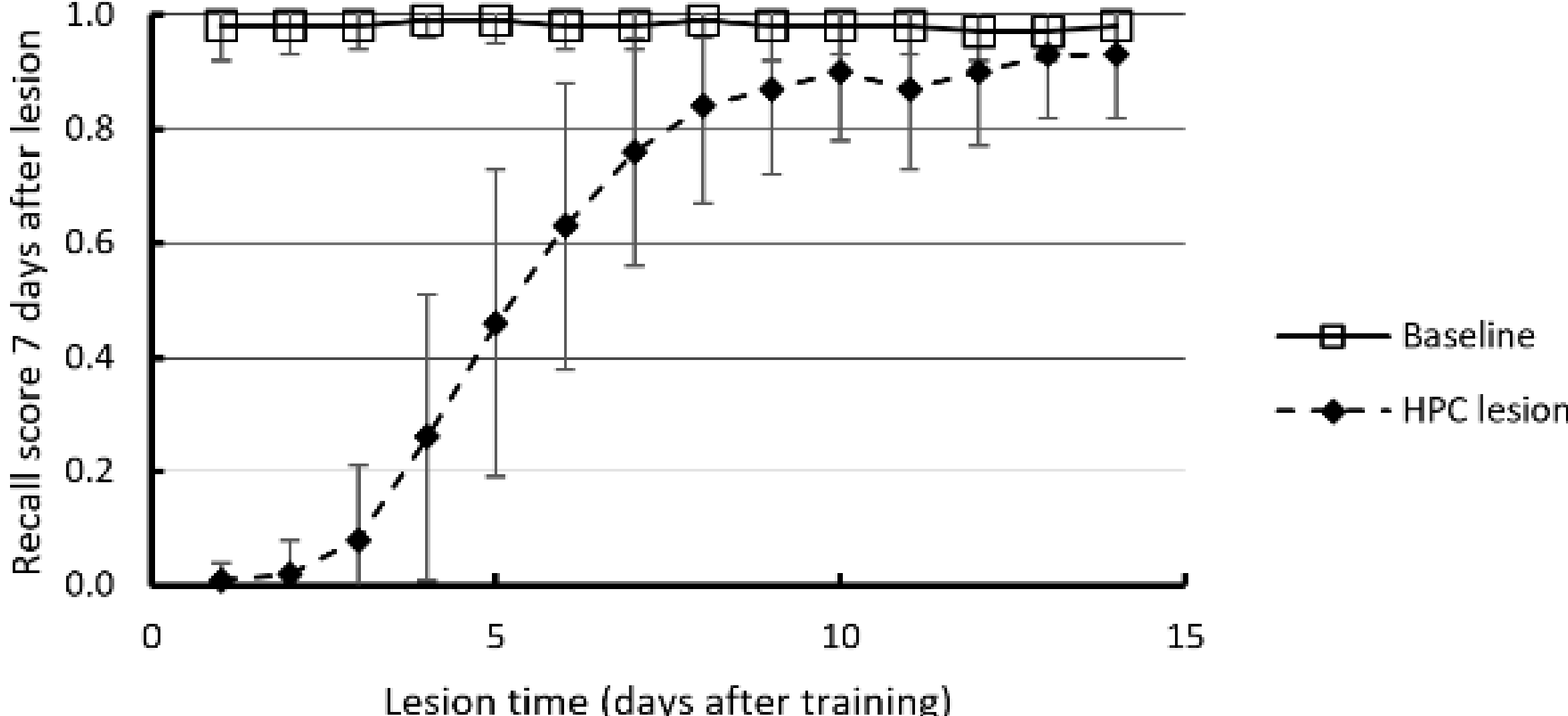

Figure 3: Consolidation window. Simulated HPC lesions produce severe impairment when performed shortly after training, but not later. Recall tests are performed 7 days after lesioning, e.g. the data point corresponding to lesions performed 5 days after conditioning reflects recall tests executed on day 12. This simulates delays used in behavioral experiments to allow the animals to recover from surgery before testing. The baseline data points reflect recall tests at the corresponding time points, but without preceding lesions.

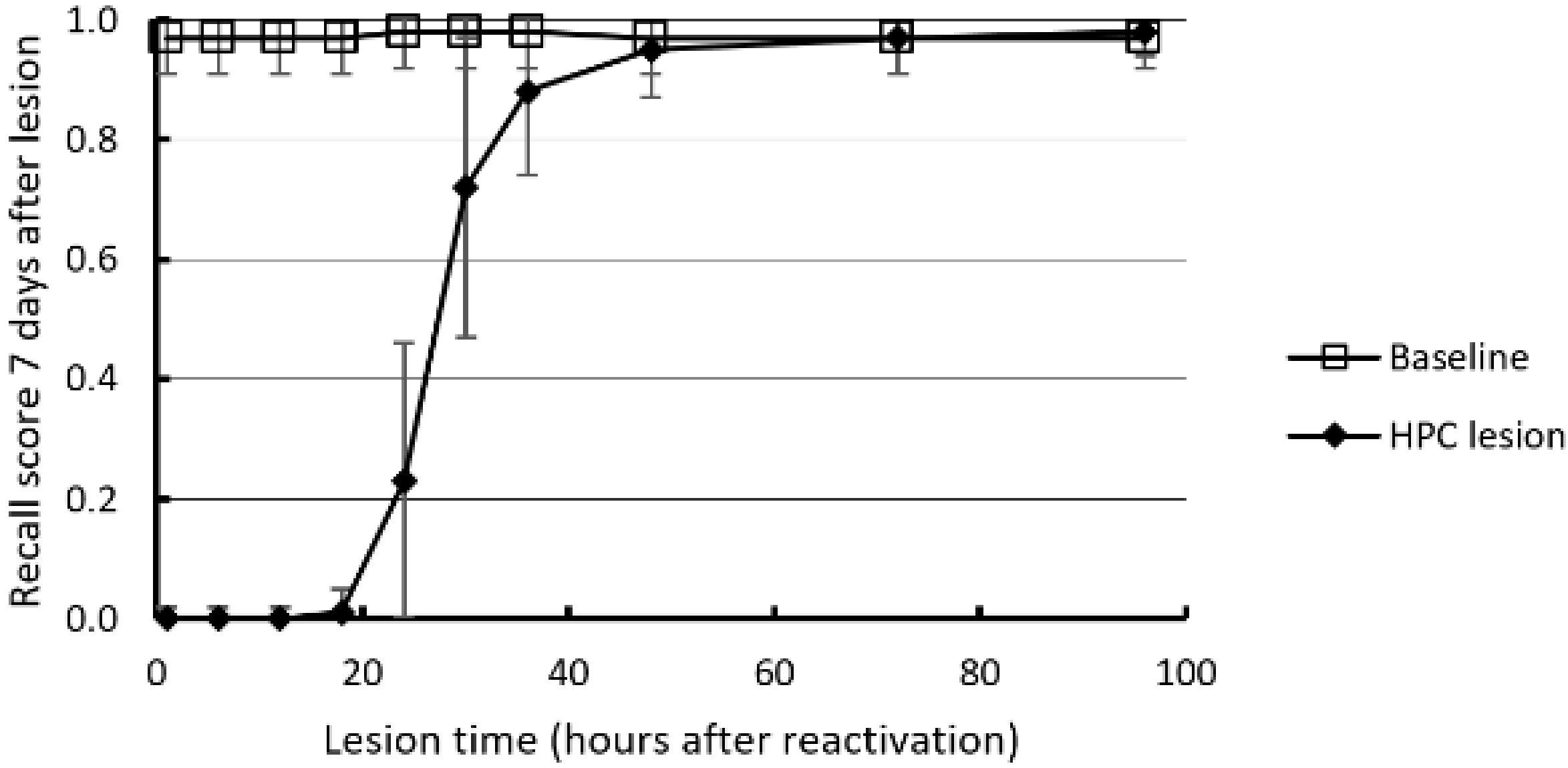

Figure 4: Reconsolidation window: Simulated HPC lesions produce severe impairment when performed shortly after reactivation, but not later. Recall tests are performed 7 days after lesioning, i.e. the data points corresponding to lesions performed 0-23 hours after reactivation reflect recall tests executed on day 7 after reactivation, etc. The baseline data points correspond to recall tests at the corresponding time points, but without preceding lesions.

**Effect of systemic PSI infusion before training.** Systemic PSI infusion before conditioning impairs formation of long-term memory but does not impair short-term memory, see Figure 5. This simulation reproduces finding 3 of Table 1.

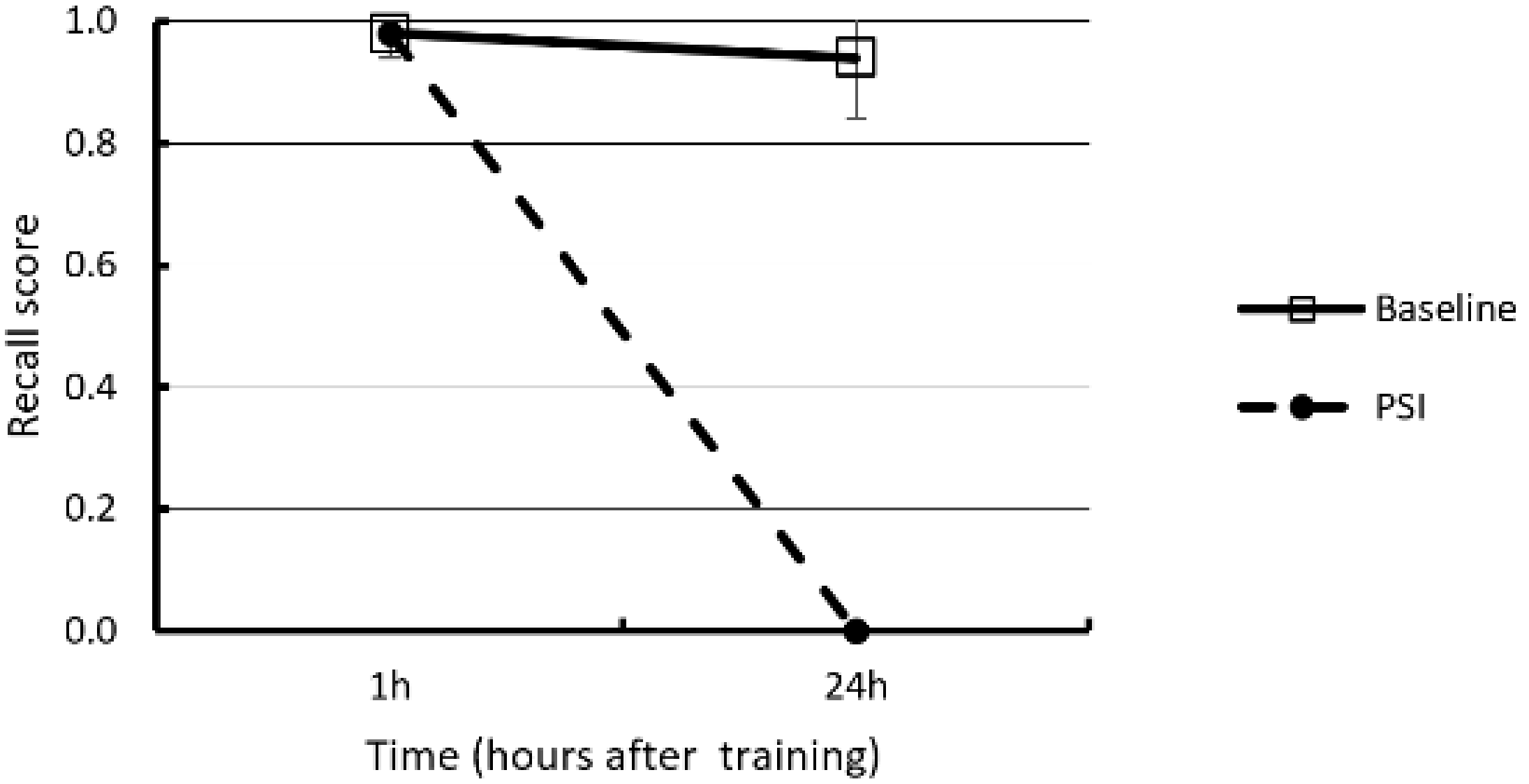

Figure 5: The effect of systemic PSI infusion before training. Short-term memory (1h) is unaffected, but long-term memory (24h) is severely impaired.

**Effect of PSI on consolidated memory.** Systemic PSI infusion during maintenance, i.e. after completed systems consolidation, does not impair subsequent memory retrieval, see Figure 6. This simulation reproduces finding 4 of Table 1.

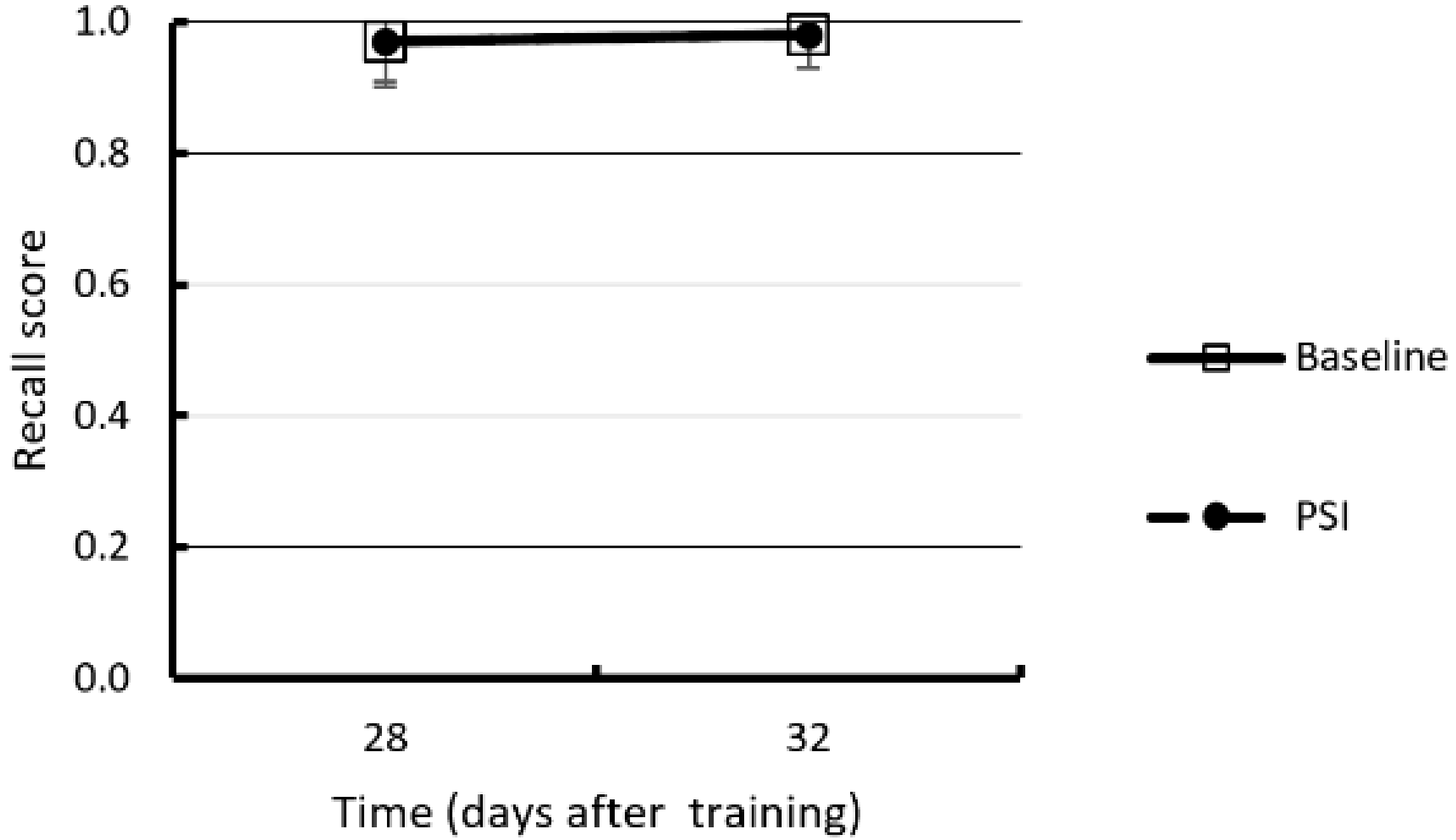

Figure 6: Effect of PSI on consolidated memory. Systemic PSI infusion is simulated on day 30 after conditioning. The diagram shows recall performance before and after the intervention. PSI infusion in the maintenance phase does not affect recall performance.

**PSI infusion in HPC after reactivation.** Post-reactivation PSI infusion in HPC does not cause immediate memory loss. Rather, the recall impairment develops over a period of more than 4 but less than 48 hours (Debiec et al., 2002), see Figure 7. This simulation reproduces findings 8 and 9 of Table 1.

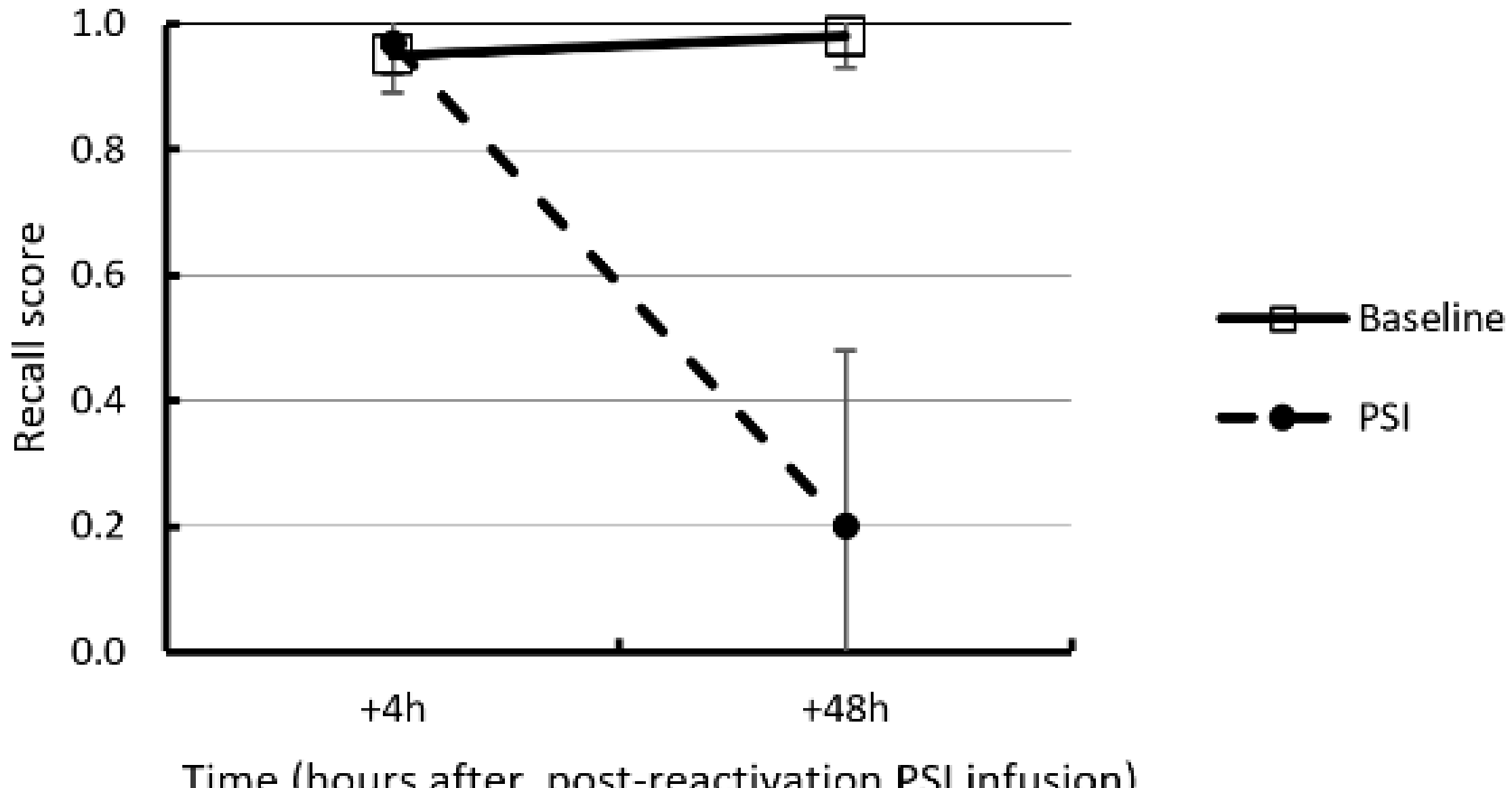

Figure 7: PSI infusion in HPC after reactivation. PSI is infused immediately after reactivation; recall tests are performed four hours later or 48 hours later. The impairment only manifests at the later time point.

**HPC/ACC-dependence for recall.** Consolidation transforms memories from being HPC-dependent to being ACC-dependent for recall, see Figure 8. These simulations reproduce findings 1 and 2 of Table 1.

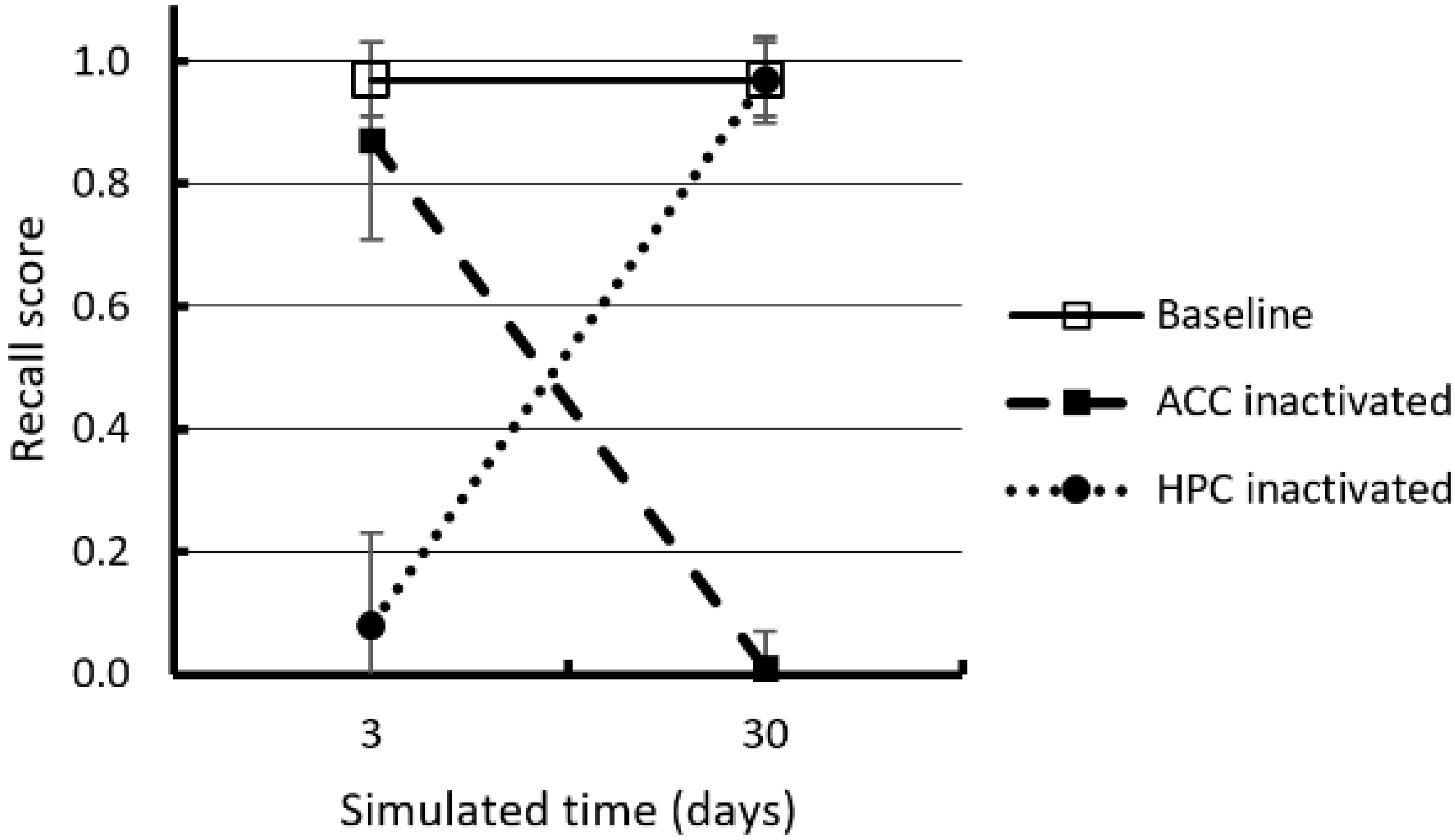

Figure 8: HPC/ACC-dependence for recall. HPC inactivation impairs recall 3 days after training, but not at 30 days. ACC inactivation does not affect recall 3 days after training, but causes severe impairment at 30 days.

**Temporary ACC-independence after reactivation.** Reactivation creates a transient HPC linkage which temporarily returns the memory to ACC-independence, see Figure 9. These simulations reproduce findings 12 and 13 of Table 1.

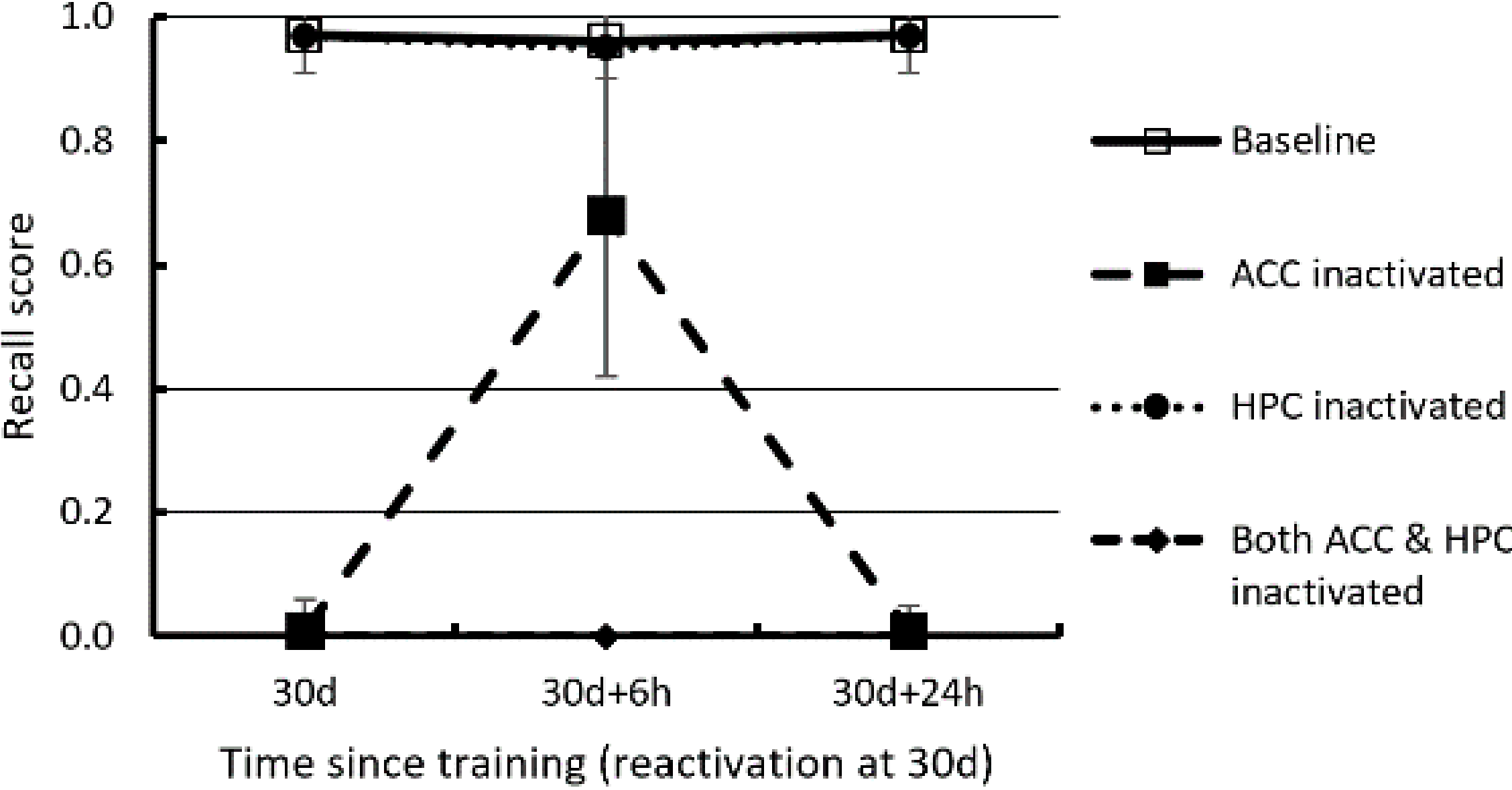

Figure 9: Temporary ACC-independence after reactivation. Before reactivation ACC inactivation severely impairs recall of a consolidated memory. Six hours after reactivation neither ACC inactivation nor HPC activation produces significant impairment. At 24h after reactivation, ACC dependence has returned.

## Discussion

We have presented an artificial neural network model of systems memory consolidation and reconsolidation that accounts for a broad range of findings from the empirical literature, including those from studies employing hippocampal lesions as well as ones using reversible inactivation of the hippocampus or anterior cingulate cortex. At the core of the model is a new connection design in which variable stability arises from simulation of receptor exchanges that have been observed in glutamatergic synapses.

**Reconsolidation**

It is worth noting that although the term “reconsolidation” suggests a recapitulation of consolidation, the model reflects our view that the two processes are quite different. Whereas systems consolidation is a gradual strengthening of the intra-neocortical connections that link together the components of a memory, systems reconsolidation consists in the restoration of stability in such synapses following reactivation-induced destabilization. HPC lesions in the “consolidation window” (the first couple of weeks after training) and in the “reconsolidation window” (the first day or two after reactivation) both result in memory deficits – but for different reasons: HPC lesions in the consolidation window prematurely interrupts the hippocampal replay that drives establishment and strengthening of neocortical linkages, leaving a weak memory trace there. Loss of HPC in the reconsolidation window, in contrast, deprives neocortical neurons of the HPC stimulation that is needed to restabilize synapses that have been destabilized by reactivation-induced AMPAR exchange. Without such stimulation, the destabilized neocortical trace decays.

Our model is the first to demonstrate results analogous to both systems consolidation and systems reconsolidation in an artificial neural network, and also to reproduce the effects of a number of experimental interventions described in the empirical literature. This was made possible by the use of a connection model with a plasticity mechanism inspired by biological synapses.

**Limitations**

Although the model is capable of reproducing a number of empirical results, it should be kept in mind that it is a restricted model with many simplifying assumptions. Thus, while the model only includes synaptic potentiation as a substrate for the creation of memory traces in systems consolidation and reconsolidation, it seems likely that other mechanisms are also at play, perhaps including axonal growth, synaptogenesis, and possibly neurogenesis (Frankland & Bontempi, 2005). Similarly, the simple receptor trafficking in the network connections is a crude model of synaptic potentiation, which is known to involve a large number of molecular processes, most of which are not well understood (Asok et al., 2019). We chose to model AMPAR trafficking because of the strong evidence of its importance for induction and maintenance of synaptic potentiation, as discussed in the introduction. Despite its limitations, it is interesting to find that a model as simple as this is able to capture a significant set of diverse findings from the consolidation/reconsolidation literature.

**Predictions**

In closing, we list a number of predictions that have been derived from the model and may be used to test its validity in future experiments.

1. In the model, post-reactivation susceptibility to HPC lesion is due to the CI-AMPAR to CP-AMPAR exchange that reactivation triggers in the consolidated connections in the ACC: subsequent CI-AMPAR restoration depends on HPC activity. A prediction that can be drawn from the model is therefore, that if reactivation is prevented from triggering AMPA receptor exchange in the ACC, then HPC lesion in the reconsolidation window will not impair recall. This could be tested by infusing a drug like GluA2$_{3Y}$ into the ACC before reactivation. GluA2$_{3Y}$ is a synthetic peptide that prevents endocytosis (removal) of CI-AMPARs from the synapse. In contrast, GluA2$_{3Y}$ should not be able to prevent the amnestic effect of hippocampal lesion during the consolidation window, because in this case the impairment is not due to depotentiation but to interrupted consolidation.

2. The amnestic effect of post-reactivation PSI infusion into HPC is also abolished if destabilization of ACC connections is blocked. The reasoning is the same as above: If the ACC connections are not destabilized, then they do not need to be restabilized, and thus depriving them of hippocampal stimulation should have no amnestic effect. So the prediction is that $GluA2_{3Y}$ infusion in ACC before reactivation would abolish the amnestic effect of post-reactivation PSI infusion in hippocampus.
3. Destabilization can also be prevented by selectively inhibiting GluN2B-containing NMDA receptors (Ben Mamou et al., 2006; Milton et al., 2013), cannabinoid receptor 1, or L-type voltage gated calcium channels (R. Kim et al., 2011). The model predicts that inhibiting any of these receptors or channels in ACC during reactivation would reduce or eliminate the amnestic effect of post-reactivation hippocampal lesion or PSI injection.
4. The model attributes the short duration of post-reactivation ACC-independence to accelerated depotentiation of linkage synapses in hippocampus. Blocking AMPAR endocytosis in hippocampus, e.g. by local $GluA2_{3Y}$ infusion, should extend the durability of reactivation-induced hippocampal memory traces, and therefore lengthen the duration of post-reactivation ACC-independence.
5. In the model, post-reactivation hippocampal lesion or PSI infusion causes memory impairments by depriving neocortical linkage synapses of the stimulation they need to restabilize. If this is correct, then other means of blocking such stimulation should have the same amnestic effect. For example, we predict that prolonged (several days) reversible inactivation of hippocampus after reactivation will impair subsequent recall of the reactivated memory.

6. In the model, all that is needed for restabilization of a reactivated memory trace is repeated activation of its ACC linkage. This suggests that it may be possible to compensate for an inactivated hippocampus (as in prediction 5) by triggering reactivations externally, i.e. by reminders. This idea is related to a result reported by Lehmann et al. (2009), where repeated conditioning sessions were shown to significantly speed up the development of hippocampus-independence.

## Data availability statement

All computer program files are available from the ModelDB database, accession number TBD (http://modeldb.yale.edu/TBD).

---